\documentclass{ejmhack}

\usepackage{hyperref}
\usepackage{amsmath,amssymb}
\usepackage{graphicx}
\usepackage{todonotes}

\let\realverbatim=\verbatim
\let\realendverbatim=\endverbatim
\renewcommand\verbatim{\par\addvspace{6pt plus 2pt minus 1pt}\realverbatim}
\renewcommand\endverbatim{\realendverbatim\addvspace{6pt plus 2pt minus 1pt}}

\ifprodtf \else
  \checkfont{eurm10}
  \iffontfound
    \IfFileExists{upmath.sty}
      {\typeout{^^JFound AMS Euler Roman fonts on the system,
                   using the 'upmath' package.^^J}%
       \usepackage{upmath}}
      {\typeout{^^JFound AMS Euler Roman fonts on the system, but you
                   don't seem to have the}%
       \typeout{'upmath' package installed. EJM.cls can take advantage
                 of these fonts,^^Jif you use 'upmath' package.^^J}%
      }
  \else
  \fi
\fi

\ifprodtf \else
  \checkfont{msam10}
  \iffontfound
    \IfFileExists{amssymb.sty}
      {\typeout{^^JFound AMS Symbol fonts on the system, using the
                'amssymb' package.^^J}%
       \usepackage{amssymb}%
         \let\leq=\leqslant
         \let\geq=\geqslant
      }{}
  \fi
\fi

\ifprodtf \else
  \IfFileExists{amsbsy.sty}
    {\typeout{^^JFound the 'amsbsy' package on the system, using it.^^J}%
     \usepackage{amsbsy}}
    {}
\fi

\newsavebox{\astrutbox}
\sbox{\astrutbox}{\rule[-5pt]{0pt}{20pt}}

\newdefinition{definition}[theorem]{Definition}

\renewcommand{\d}{{\rm d}}
\newcommand{\e}{{\rm e}}

\newcommand{\FD}[2]{\frac{\d #1}{\d #2}}
\renewcommand{\vec}[1]{\mathbf{#1}}

\DeclareMathSymbol{\ZSet}{\mathalpha}{AMSb}{"5A}
\DeclareMathSymbol{\RSet}{\mathalpha}{AMSb}{"52}
\DeclareMathSymbol{\CSet}{\mathalpha}{AMSb}{"43}

\title[Networks of piecewise linear neural mass models]{Networks of piecewise linear neural mass models}

\author[S. Coombes et al.]{%
  S.\ns C\ls O\ls O\ls M\ls B\ls E\ls S,\ns
  Y.\ns M.\ns L\ls A\ls I,\ns
  M. \ns \c{S}\ls A\ls Y\ls L\ls I,\ns
\and
  R.\ns T\ls H\ls U\ls L.\ns
}

\affiliation{Centre for Mathematical Medicine and Biology, School of Mathematical Sciences, University of Nottingham,
Nottingham, NG7 2RD, UK}

\date{\today}
\pubyear{2016}
\volume{000}

\pagerange{\pageref{firstpage}--\pageref{lastpage}}

\begin{document}

\label{firstpage}


\maketitle

\begin{abstract}
Neural mass models are ubiquitous in large scale brain modelling.  At the node level they are written in terms of a set of ordinary differential equations with a nonlinearity that is typically a sigmoidal shape.  Using structural data from brain atlases they may be connected into a network to investigate the emergence of functional dynamic states, such as synchrony.  With the simple restriction of the classic sigmoidal nonlinearity to a piecewise linear caricature we show that the famous Wilson-Cowan neural mass model can be explicitly analysed at both the node and network level.  The construction of periodic orbits at the node level is achieved by patching together matrix exponential solutions, and stability is determined using Floquet theory.  For networks with interactions described by circulant matrices, we show that the stability of the synchronous state can be determined in terms of a low-dimensional Floquet problem parameterised by the eigenvalues of the interaction matrix.  Moreover, this network Floquet problem is readily solved using linear algebra, to predict the onset of spatio-temporal network patterns arising from a synchronous instability.  We further consider the case of a discontinuous choice for the node nonlinearity, namely the replacement of the sigmoid by a Heaviside nonlinearity.  This gives rise to a \textit{continuous-time switching network}.
At the node level this allows for the existence of unstable sliding periodic orbits, which we explicitly construct.  The stability of a periodic orbit is now treated with a modification of Floquet theory to treat the evolution of small perturbations through \textit{switching manifolds} via the use of \textit{saltation matrices}.  At the network level the stability analysis of the synchronous state is considerably more challenging.  Here we report on the use of ideas originally developed for the study of Glass networks to treat the stability of periodic network states in neural mass models with discontinuous interactions.
\end{abstract}


\section{\label{sec:introduction}Introduction}

The Wilson-Cowan model \cite{Wilson72,Wilson73} is one of the most well-known neural mass models for modelling the activity of cortex, and for a historical perspective see \cite{Cowan04}.  Neural mass models generate brain rhythms using the notion of population firing rates, aiming to side-step the need for large scale simulations of more realistic networks of spiking neurons.  Although they are not derived from detailed conductance based models they can be motivated by a number of phenomenological arguments \cite{Coombes2014}, and typically take the form of systems of nonlinear ordinary differential equations (ODEs).  The Wilson-Cowan neural mass model describes the dynamics of two interacting populations of neurons, one of which is excitatory and the other inhibitory.  Interactions are mediated between the populations with the use of a nonlinear sigmoidal firing rate function.
In its most simple incarnation it consists of two nonlinear ODEs, and as such has been widely studied using techniques from phase-plane analysis and numerical bifurcation theory.  At the network level the model can either be posed on a graph or a continuous space, and since the 1970s there has been a large amount of attention devoted to the analysis of these models and their application in neuroscience \cite{Destexhe2009}.  Recent examples of their use include reconciling information from anatomical and functional data \cite{Woolrich2013}, understanding \textit{phase-amplitude} coupling (whereby the amplitude of a higher frequency brain rhythm is modulated by the phase of lower frequency activity) \cite{Onslow2014}, modelling epilepsy \cite{Meijer2015}, and understanding the emergence of cortical resonant frequencies \cite{Lea-Carnall2016}.
Indeed there are many variants of the Wilson-Cowan neural mass model now in use for interpreting neuroimaging data \cite{Valdes-Sosa2009}, including those of Zetterberg \textit{et al}. \cite{Zetterberg1978}, Jansen and Rit \cite{Jansen95}, and Liley \textit{et al}. \cite{Liley02}.  Neural mass models are a key component of the Virtual Brain project that aims to deliver the first simulation of the human brain based on individual large-scale connectivity \cite{Sanz-Leon2015}.   Such large-scale brain network models are especially relevant to understanding resting state networks \cite{Breakspear2017}, whereby different regions of the brain's sensorimotor system oscillate slowly and synchronously in the absence of any explicit task.

However, at heart it is well to note that from a mathematical modelling perspective all neural mass models to date are essentially low dimensional coupled ODEs with a sigmoidal firing rate nonlinearity, exemplified by the Wilson-Cowan model.  Using extensions of the techniques originally developed by Amari \cite{Amari77}, the continuum or \textit{neural field} \cite{Coombes2014} Wilson-Cowan model has been analysed when the choice of this firing rate nonlinearity is a Heaviside function.  This has been possible because of a \textit{smoothing} of the firing rate with a spatial kernel representing anatomical connectivity. However, when posed on a graph, representing a network of interacting neural populations, no such smoothing arises.  Surprisingly there are hardly any mathematical results for such networks, as opposed to their continuum counterparts for which there are now a plethora ranging from the properties of localised states through to travelling waves, as reviewed in \cite{Coombes05}. This discrepancy is really a reflection of the fact that there are many more techniques for studying smooth dynamical systems as opposed to nonsmooth.  However, the body of mathematical work in this area is rapidly growing, driven in part by its importance to engineering \cite{diBernardo2008,Colombo2012}.  Given their relevance to large scale brain dynamics it is highly desirable to develop mathematical techniques for the analysis of Wilson-Cowan style neural mass models at the network level.  Here we advocate for the replacement of smooth sigmoidal nonlinearities in neural mass models by more tractable functions, including piecewise linear (PWL) and piecewise constant functions.  A PWL continuous choice has been used in several previous studies, including those of Hansel and Sompolinsky \cite{Hansel1998}, and Kilpatrick and Bressloff \cite{Kilpatrick2010}, whilst the discontinuous Heaviside (piecewise constant) choice has proven especially popular since the seminal work of Amari \cite{Amari77}.  In these instances this has facilitated the construction of certain types of localised states in continuum neural field models.  However, in a discrete neural network context there is a major mathematical difference in the analysis of network states for the case of continuous vs. discontinuous firing rates.  As well as introducing a simple methodology to treat the construction of periodic orbits in idealised Wilson-Cowan networks, this is one of the major topics we wish to address in this paper.

First in \S \ref{sec:WilsonCowan} we introduce the model for an isolated Wilson-Cowan node with a PWL firing rate. 
The description of dynamical states with reference to \textit{switching manifolds} becomes very useful.  
We show how matrix exponentials can be used to patch together a periodic orbit, and that Floquet theory simplifies considerably to yield explicit formulas for determining solution stability.  Next in \S \ref{sec:WilsonCowanNetwork} we consider a network of PWL Wilson-Cowan nodes, with nodes arranged along a ring with distance-dependent interactions.  This particular choice of coupling guarantees the existence of the synchronous state.  We then develop a linear stability analysis of this state and show that this leads to a tractable variational problem of a very similar type to that for the single node, albeit now parameterised by the eigenvalues of the connectivity matrix.  We use this to determine instabilities that can lead to the formation of spatio-temporal network patterns.  Next in \S \ref{sec:Heaviside} we consider the case that the firing rate is a Heaviside function, for which the techniques developed for studying PWL systems break down. Once again periodic orbits can be constructed using matrix exponentials, although standard Floquet theory must be now augmented to cope with the evolution of linearised perturbations through the switching manifolds.  This is most readily achieved with the use of saltation matrices that have commonly been used for the study of nonsmooth mechanical systems \cite{Leine2004}.  However, at the network level the stability of the synchronous state is much harder to determine than for the continuous model.  Here we show that ideas from the study of Glass networks developed by Edwards \cite{Edwards2000} are particularly useful, and that stability is strongly influenced by the temporal order in which network components cross switching manifolds, and that this in turn is determined by the choice of initial perturbation.  Finally in \S \ref{sec:conclusion} we conclude with an overview of the new results about synchrony in networks of neural mass models, and discuss the natural extension of this work to treat non-synchronous states.

\section{\label{sec:WilsonCowan}The Wilson-Cowan model and a piecewise linear reduction}

\begin{figure}[htbp]
\begin{center}
\includegraphics[width=8cm]{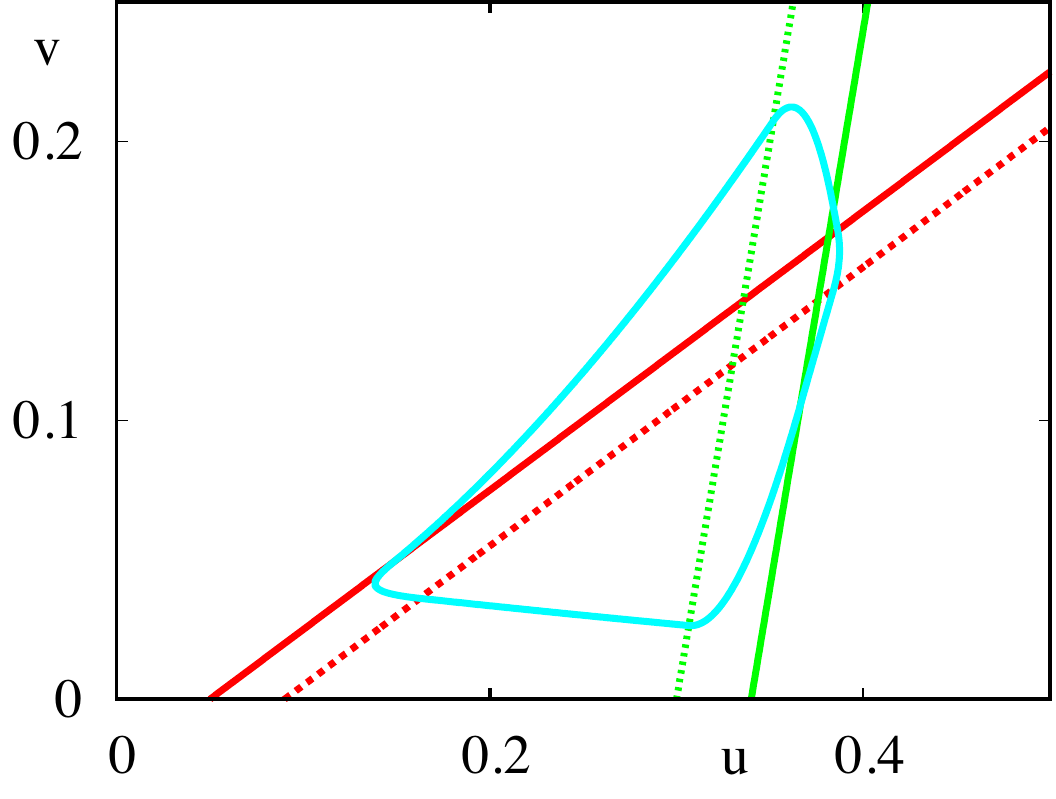}
\caption{Phase plane for the Wilson-Cowan network with a PWL firing rate, showing a stable periodic orbit (light blue).
Parameters:  $\epsilon=0.04$, $\tau=0.6$, $I_u=-0.05$, $I_v=-0.3$, $w^{uu}=1$, $w^{vu}=2$, $w^{uv}=1$, and $w^{vv}=0.25$.
The straight lines in red and green show the \textit{switching manifolds}, where $I_u+ w^{uu}u-w^{vu}v=0,\epsilon$ and $I_v + w^{uv} u - w^{vv}v=0,\epsilon$ respectively.
\label{Fig:Orbit}
}
\end{center}
\end{figure}

\begin{figure}[htbp]
\begin{center}
\includegraphics[width=8cm]{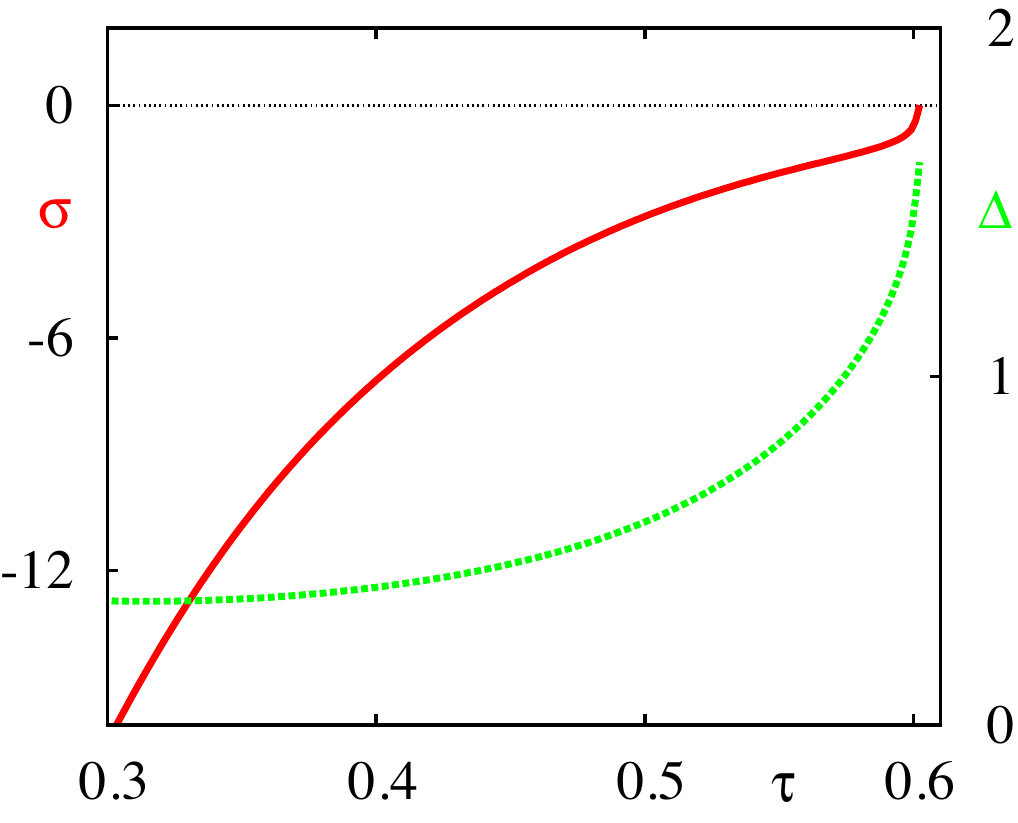}
\caption{A plot of the non-trivial Floquet exponent for the PWL Wilson-Cowan model (left axis), as a function of the relative time-scale $\tau$, with the period of the orbit also shown (right axis).  Parameters as in Fig.~\ref{Fig:Orbit}. Periodic orbits emerge via a supercritical Hopf bifurcation as $\tau$ increases through $\tau_{\text{Hopf}}=(w^{vv}+\epsilon)/(w^{uu}-\epsilon) \sim 0.3$.
We see that the branch of periodic orbits shown is stable, with stability decreasing to zero as the solution is lost with increasing $\tau$.   This loss of existence occurs because of a \textit{grazing} bifurcation (coincident with a saddle-node bifurcation of periodic orbits) at $\tau_{\text{graze}} \sim 0.6$ whereby part of the trajectory develops a point of inflection on the switching manifold $v=(I_u +w^{uu}u)/w^{vu}$ (red solid line in Fig.~\ref{Fig:Orbit}), such that beyond bifurcation the trajectory does not cross the switching manifold and instead is attracted to the stable fixed point at $(u,v)=(0,0)$.
\label{Fig:sigma}
}
\end{center}
\end{figure}

For their activity-based neural mass model Wilson and Cowan \cite{Wilson72,Wilson73} distinguished between excitatory and inhibitory sub-populations.
This seminal (space-clamped) model can be written succinctly in terms of the pair of coupled differential equations:
\begin{equation}
\FD{u}{t}  = -u + F(I_u+ w^{uu}u-w^{vu}v), \qquad \tau \FD{v}{t} = - v + F(I_v + w^{uv} u - w^{vv}v) ,
\label{WC}
\end{equation}
Here $u=u(t)$ is a temporal coarse-grained variable describing the proportion of excitatory cells firing per unit time at the instant $t$.  Similarly the variable $v$ represents the activity of an inhibitory population of cells. The constants $w^{\alpha \beta}$, $\alpha,\beta \in \{u,v\}$, describe the weight of all synapses from the $\alpha$th population to cells of the $\beta$th population, and $\tau$ is a relative time-scale.  The nonlinear function $F$ describes the expected proportion of neurons in population $\alpha$ receiving at least threshold excitation per unit time, and is often taken to have a sigmoidal form.  Here the terms $I_\alpha$ represent external inputs (that could be time varying).
For a historical perspective on the Wilson-Cowan model see \cite{Destexhe2009}, and for a more recent reflection by Cowan see \cite{Cowan2016}.
To reduce the model to a mathematically tractable form we consider the choice of a PWL firing rate function given by
\begin{equation}
F(x) = \begin{cases}
0 & x \leq 0 \\
\epsilon^{-1}x & 0 < x < \epsilon \\
1 & x \geq \epsilon
\end{cases} .
\label{pwlrate}
\end{equation}
For appropriate choices of parameters the Wilson-Cowan model, with the firing rate given by (\ref{pwlrate}), can support stable oscillations.  An example is shown in Fig.~\ref{Fig:Orbit}, where we also plot the four \textit{switching manifolds} defined by the condition that arguments to the function $F$ in (\ref{WC}) take on the values zero and $\epsilon$.  
Away from the switching manifolds the dynamics governing the evolution of trajectories is linear, and may be constructed using matrix exponentials.  
To simplify further analysis it is first convenient to introduce new variables $(U,V)$ such that $u=(w^{vu}(V-I_v)-w^{vv}(U-I_u))/|W|$, where $|W|=\det W$, and $v=(w^{uu}(V-I_v)-w^{uv}(U-I_u))/|W|$, as well as the matrices
\begin{equation}
W=\begin{bmatrix}
w^{uu} & - w^{vu} \\
w^{uv} & - w^{vv}
\end{bmatrix},
\qquad
J=\begin{bmatrix}
1 & 0 \\
0 & 1/\tau
\end{bmatrix} \qquad
A = - W J W^{-1} .
\label{J}
\end{equation}
With these choices (\ref{WC}) transforms to
\begin{equation}
\FD{}{t} \begin{bmatrix} U \\ V \end{bmatrix} = A \begin{bmatrix} U -  I_u \\ V  -  I_v\end{bmatrix} + W J   \begin{bmatrix} F(U) \\ F(V) \end{bmatrix}
\label{WC1} .
\end{equation}
In the representation (\ref{WC1}) we see that the four switching manifolds are simply defined by $U=0$, $U=\epsilon$, $V=0$, and $V=\epsilon$.
The periodic orbit shown in Fig.~\ref{Fig:Orbit} (encircling an unstable fixed point) crosses each of these manifolds twice, so that the periodic trajectory is naturally decomposed into eight separate pieces.  On each piece we shall denote the \textit{time-of-flight} for a trajectory to travel from one switching manifold to another by $\Delta_i$, $i=1,\ldots,8$, so that the period of the orbit is given by $\Delta = \sum_{i=1}^8 \Delta_i$.  As an explicit example of how to construct a trajectory between two switching manifolds, consider the region
where $0 \leq U \leq \epsilon$ and $V<0$. In this case the solution of (\ref{WC1}) is given by 
\begin{equation}
\begin{bmatrix} U(t) \\ V(t) \end{bmatrix}=\e^{A_+(\epsilon)t}\begin{bmatrix} U(0) \\ V(0) \end{bmatrix}-{A}_+^{-1}(\epsilon)\left( \e^{A_+(\epsilon)t}-I_2 \right) A \begin{bmatrix} I_u \\ I_v  \end{bmatrix}, \qquad t \geq 0, 
\end{equation}
where 
\begin{equation}
A_+(\epsilon)=\left( A + \epsilon^{-1}W J   \begin{bmatrix} 1 & 0 \\ 0 & 0 \end{bmatrix}\right) .
\end{equation}
It is a simple matter to write down the trajectories in each of the remaining regions of phase space visited by a periodic orbit.  We may then use these matrix exponential formulas to \textit{patch} together solutions, setting the origin of time in each region such that \textit{initial} data in one region comes from \textit{final} data from a trajectory in a neighbouring region.  We shall denote the periodic orbit by $(\overline{U},\overline{V})$ such that $(\overline{U}(t),\overline{V}(t))=(\overline{U}(t+\Delta),\overline{V}(t+\Delta))$.  If we consider initial data with $(\overline{U}(0),\overline{V}(0))=(U_0,0)$ then the eight times-of-flight and the unknown $U_0$
are determined self-consistently by the nine equations $\overline{V}(\Delta_1)=\epsilon$, $\overline{U}(\Delta_2)=\epsilon$, $\overline{U}(\Delta_3)=0$, $\overline{V}(\Delta_4)=\epsilon$, $\overline{V}(\Delta_5)=0$, $\overline{U}(\Delta_6)=0$, $\overline{U}(\Delta_7)=\epsilon$, $\overline{V}(\Delta_8)=0$, and $\overline{U}(\Delta_8)=U_0$.  The numerical solution of this nonlinear algebraic system of equations can be used to construct periodic orbits such as the one shown in Fig.~\ref{Fig:Orbit}.  Note that the construction of periodic orbits that do not cross all of the switching manifolds can similarly be performed (requiring the simultaneous solution of fewer equations).  To determine stability we can turn directly to Floquet theory for planar systems which tells us that the non-trivial Floquet exponent is given by 
\begin{equation}
\sigma = \frac{1}{\Delta} \int_0^\Delta \operatorname{Tr} D(s) \d s,
\end{equation}
where $D(s)$ denotes the Jacobian of the system evaluated along the periodic orbit.  In general this is a hard quantity to evaluate for systems where the periodic orbit is not available in closed form.  However, for the PWL Wilson-Cowan model the Jacobian is piecewise constant and 
we have that
\begin{equation}
\sigma = \frac{1}{\Delta} \sum_{i=1}^8 \Delta_i \operatorname{Tr} A_i ,
\end{equation}
where $A_2=A_4=A_6=A_8=A$, $A_3=A_7=A_+(\epsilon)$, and $A_1=A_5=A_-(\epsilon)$, where
\begin{equation}
A_-(\epsilon)=\left( A + \epsilon^{-1}W J   \begin{bmatrix} 0 & 0 \\ 0 & 1 \end{bmatrix}\right) .
\end{equation}
Thus a periodic orbit is stable if $\sigma <0$.
In Fig.~\ref{Fig:sigma} we present a plot of $\sigma$ as a function of $\tau$, to show that the periodic solution in Fig.~\ref{Fig:Orbit} is stable.

Given the above method to construct and determine the stability of a periodic orbit, we next show how to extend this aproach to treat synchronous solutions in networks of Wilson-Cowan oscillators.

\section{\label{sec:WilsonCowanNetwork}A piecewise linear Wilson-Cowan network}

The study of coupled oscillator networks in biology, physics, and engineering is now commonplace.  Two particularly well known tools for studying patterns of phase-locked states and their instabilities are the theory of weakly coupled oscillators \cite{Kuramoto84}, and the master stability function (MSF) \cite{Pecora1998}.  The reduction of a coupled limit cycle network to a phase oscillator network has proven very useful for gaining insight into phenomena ranging from the synchronisation of flashing fireflies \cite{Ermentrout1991} to behaviours in social networks \cite{Assenza2011}, and for a recent review see \cite{Dorfler2014}.  However, there is an obvious limitation to such an approach, namely the restriction to \textit{weak} interaction (and near identical oscillators).  The MSF approach (for identical oscillators) does not require any such restriction on coupling strength, and can be used to determine the stability of the synchronous state in terms of the eigen-structure of the network connectivity matrix.  However, the numerical evolution of a system of dynamical equations, arising from a Floquet variational problem, must be performed.  Importantly the MSF approach can be combined with group theoretical techniques used in the study of symmetric dynamical systems to analyse the stability of cluster states within symmetric networks of dynamical units
\cite{Pecora2014,Sorrentino2016}.  Here we favour the MSF approach and show it simplifies considerably for a PWL choice of firing rate function.  This allows us to improve upon previous mathematical studies of Wilson-Cowan networks, such as those by Campbell and Wang \cite{Campbell1996} (who treated networks with nearest neighbour coupling and established the condition for synchrony), Ueta and Chen \cite{Ueta2003} (who performed a numerical bifurcation analysis for small networks), and Ahmadizadeh \textit{et al}. \cite{Ahmadizadeh2015} (who used perturbation techniques and numerics to study synchrony in networks with diffusive coupling).

We now consider a network of Wilson-Cowan nodes given by
\begin{align}
\FD{u_i}{t}  &= -u_i + F \left (I_u+ \sum_{j=1}^N \mathcal{W}^{uu}_{ij}u_j-\sum_{j=1}^N \mathcal{W}^{vu}_{ij}v_j \right ), \\
\tau \FD{v_i}{t} &= - v_i + F \left (I_v + \sum_{j=1}^N \mathcal{W}^{uv}_{ij} u_j - \sum_{j=1}^N \mathcal{W}^{vv}_{ij}v_j \right) , \qquad i=1,\ldots,N,
\label{WCnetwork}
\end{align}
subject to the constraints $\sum_{j=1}^N \mathcal{W}^{uu}_{ij} = w^{uu}$, $\sum_{j=1}^N \mathcal{W}^{vu}_{ij}=w^{vu}$, $\sum_{j=1}^N \mathcal{W}^{uv}_{ij}=w^{uv}$, and $\sum_{j=1}^N\mathcal{W}^{vv}_{ij}=w^{vv}$ for all $i$.  These \textit{row-sum} constraints are natural for networks arranged on a ring, and guarantee the existence of a synchronous orbit $(u_i(t), v_i(t)) = (u(t), v(t))$ for all $i=1,\ldots,N$, where $(u(t), v(t))$ is given by the solution of (\ref{WC}).  

It is now convenient to introduce a vector notation with $X=(u_1, v_1, u_2, v_2, \ldots , u_N, v_N) \in \RSet^{2N}$ and consider a change of variables $Y = \mathcal{W} X + C$, where $C = \vec{1}_N \otimes (I_u, I_v)$, and
\begin{equation}
\mathcal{W} = 
\mathcal{W}^{uu} \otimes \begin{bmatrix} 1 & 0 \\ 0 & 0	\end{bmatrix}
- \mathcal{W}^{vu} \otimes \begin{bmatrix} 0 & 1 \\ 0 & 0	\end{bmatrix}
+ \mathcal{W}^{uv} \otimes \begin{bmatrix} 0 & 0 \\ 1 & 0	\end{bmatrix}
- \mathcal{W}^{vv} \otimes \begin{bmatrix} 0 & 0 \\ 0 & 1	\end{bmatrix}
.
\end{equation}
Here the symbol $\otimes$ denotes the usual tensor product for matrices, and $\vec{1}_N$ is an $N$-dimensional vector with all entries equal to unity.
This means that the switching manifolds can be succinctly described by $Y_i=0$ and $Y_i=\epsilon$, and the dynamics takes the form
\begin{equation}
\label{eq:WCN_dY}
\FD{}{t} Y = \mathcal{A} (Y- C) + \mathcal{W} \mathcal{J} F(Y),
\end{equation}
where 
\begin{equation}
\mathcal{J}=I_N \otimes J , \qquad
\mathcal{A} = - \mathcal{W} \mathcal{J} \mathcal{W}^{-1} ,
\end{equation}
where $J$ is given by (\ref{J}) and $I_N$ is the $N \times N$ identity matrix.
If we denote the synchronous solution by $\overline{Y}(t)= (\overline{U}(t), \overline{V}(t), \overline{U}(t), \overline{V}(t),  \ldots, \overline{U}(t), \overline{V}(t))$ and consider small perturbations such that $Y = \overline{Y} + \delta Y$, then these evolve according to
\begin{equation}
\FD{}{t} \delta Y = \mathcal{A} \delta Y + \mathcal{W} \mathcal{J} DF(\overline{Y}) \delta Y ,
\label{perturbN}
\end{equation}
where $DF(\overline{Y})$ is the Jacobian of $F$ evaluated along the periodic orbit.

Given the constraints on the matrices $\mathcal{W}^{\alpha \beta}$, with $\alpha, \beta \in \{u,v\}$ it is natural to take these to be circulant matrices with $\mathcal{W}^{\alpha \beta}_{ij} = \mathcal{W}^{\alpha \beta}_{|i-j|}$.  In this case the normalised eigenvectors of $\mathcal{W}^{\alpha \beta}$ are given by $e_p = (1,\omega_p, \omega_p^2, \ldots, \omega_p^{N-1})/\sqrt{N}$, where $p=0,\ldots, N-1$, and $\omega_p=\exp(2 \pi i p /N)$ are the $N$th roots of unity.  The corresponding complex eigenvalues are given by $\nu^{\alpha \beta} = \nu^{\alpha \beta}(p)$ where
\begin{equation}
\nu^{\alpha \beta}(p) = \sum_{\mu=0}^{N-1} \mathcal{W}^{\alpha \beta}_{\mu} \omega_p^\mu .
\end{equation}
If we introduce the matrix of eigenvectors $P=[e_0 ~e_1 ~\ldots ~e_{N-1}]$, then we have that
\begin{align}
(P \otimes I_2)^{-1}\mathcal{W} (P \otimes I_2) &=
\Lambda^{uu} \otimes \begin{bmatrix} 1 & 0 \\ 0 & 0	\end{bmatrix}
-\Lambda^{vu}  \otimes \begin{bmatrix} 0 & 1 \\ 0 & 0	\end{bmatrix}
+\Lambda^{uv}  \otimes \begin{bmatrix} 0 & 0 \\ 1 & 0	\end{bmatrix}
-\Lambda^{vv} \otimes \begin{bmatrix} 0 & 0 \\ 0 & 1	\end{bmatrix}, \nonumber \\
&= \operatorname{diag}(\Lambda(0),\Lambda(1), \ldots, \Lambda(N-1)) \equiv \Lambda ,
\end{align}
where $\Lambda^{\alpha \beta} = \operatorname{diag}(\nu^{\alpha \beta}(0),\nu^{\alpha \beta}(1), \ldots, \nu^{\alpha \beta}(N-1))$, and
\begin{equation}
\Lambda(p) = \begin{bmatrix} \nu^{uu}(p) & - \nu^{vu}(p) \\
\nu^{uv}(p) & - \nu^{vv}(p)
\end{bmatrix}, \qquad p=0,1,\ldots, N-1 .
\end{equation}
Moreover, it is easy to establish that in the above notation $(P \otimes I_2)^{-1}\mathcal{A} (P \otimes I_2) = -\Lambda (I_N \otimes J) \Lambda^{-1}$.

If we now consider perturbations of the form $\delta Z = (P \otimes I_2)^{-1}\delta Y$ then from (\ref{perturbN}) we find that the linearised dynamics is described by the system
\begin{equation}
\FD{}{t} \delta Z = \Lambda (I_N \otimes J)\left [ -\Lambda^{-1}  + (I_N \otimes D)\right ] \delta Z ,
\label{deltaZ}
\end{equation}
where $D \in \RSet^{2 \times 2}$ is the Jacobian of $(F(\overline{U}),F(\overline{V}))$, and is a piecewise constant matrix that is only non-zero if $0 < \overline{U}(t) < \epsilon$ or $0  < \overline{V}(t) < \epsilon$.
In the former case $[DF]_{11}=\epsilon^{-1}$ with all other entries zero, and in the latter case $[DF]_{22}=\epsilon^{-1}$ with all other entries zero.  
We see that (\ref{deltaZ}) has a block structure where the dynamics in each of $N$ $2 \times 2$ blocks is given by
\begin{equation}
\FD{}{t} \xi
= [A(p) + \Lambda(p) J D]\xi  , \qquad p=0,\ldots, N-1, \qquad \xi \in \RSet^2,
\label{cool}
\end{equation}
with $A(p)= - \Lambda(p)  J \Lambda^{-1}(p)$.
Thus, comparing to (\ref{WC1}), we see that the variational equation for the network is identical to that for a single Wilson-Cowan unit with $W$ replaced by $\Lambda(p)$.  We note that for $p=0$ the variational problem is identical to that for an isolated node since $\Lambda(0)=W$ (using $\nu^{\alpha \beta}(0) = \sum_{\mu=0}^{N-1} \mathcal{W}_\mu^{\alpha \beta} = w^{\alpha \beta}$).  Thus to determine the stability of the synchronous state we only have to consider a set of $N$ two dimensional variational problems.  Exploiting the fact that between switching manifolds the variational problem defined by (\ref{cool}) is time-independent we may construct a solution in a piecewise fashion from matrix exponentials and write $\xi(t) = \exp [(A(p) + \Lambda(p) J D)t] \xi(0)$.  We may then build up a perturbed trajectory over one period of oscillation in the form $\xi(\Delta)=\Gamma(p) \xi(0)$, where $\Gamma(p) \in \RSet^{2 \times 2}$ is given by
\begin{equation}
\Gamma(p)= \e^{A(p)\Delta_8}\e^{A_+(p;\epsilon)\Delta_7}\e^{A(p)\Delta_6}\e^{A_-(p;\epsilon)\Delta_5}\e^{A(p)\Delta_4}\e^{A_+(p;\epsilon)\Delta_3}\e^{A(p)\Delta_2}\e^{A_-(p;\epsilon)\Delta_1},
\label{Gamma}
\end{equation}
where
\begin{equation}
A_+(p;\epsilon)=\left( A(p) + \epsilon^{-1} \Lambda(p) J   \begin{bmatrix} 1 & 0 \\ 0 & 0 \end{bmatrix}\right) ,\quad
A_-(p;\epsilon)=\left( A(p) + \epsilon^{-1} \Lambda(p) J   \begin{bmatrix} 0 & 0 \\ 0 & 1 \end{bmatrix}\right) .
\end{equation}
Thus if a periodic orbit of an isolated Wilson-Cowan node is stable then the synchronous network solution will be stable provided all the eigenvalues of $\Gamma(p)$, for $p=0,\ldots,N-1$, lie in the unit disc (excluding the one that arises from time-translation invariance, with a value $+1$).  
For a fixed value of $p$ one of three bifurcations is possible, namely a tangent instability defined by $\det (\Gamma(p)-I_2)=0$, a period-doubling instability defined by $\det (\Gamma(p)+I_2)=0$, and a Neimark-Sacker bifurcation defined by $\det \Gamma(p)=1$.  If there is a $p=p_c$ such that one of these instabilities occurs then the excited network state will correspond to the eigenvector $\text{Re} \,e_{p_c}$.

\subsection{Example: a ring network}

\begin{figure}[htbp]
\begin{center}
\includegraphics[width=5cm]{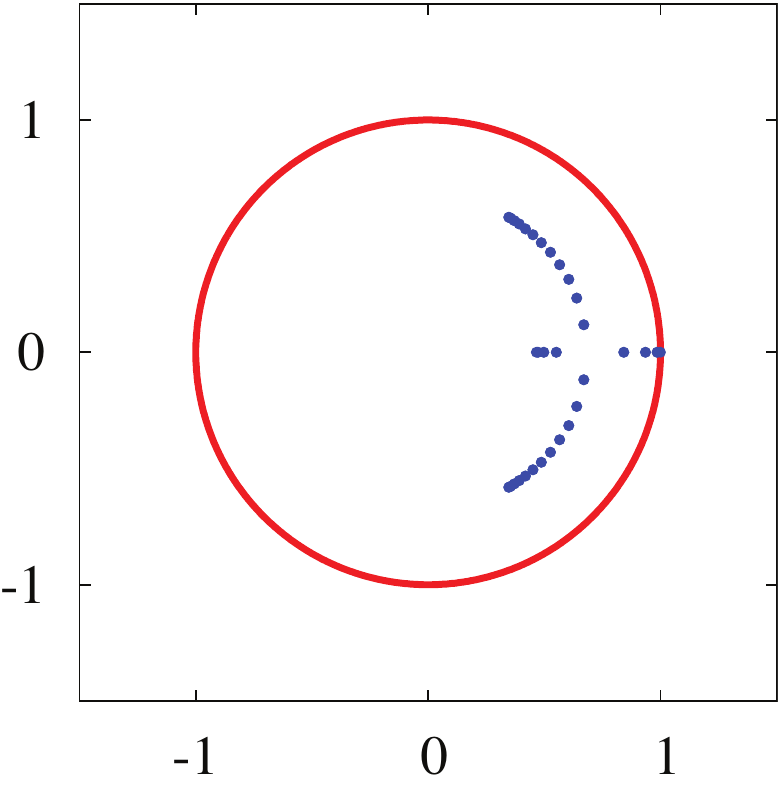} \hspace*{.75cm}
\includegraphics[width=5cm]{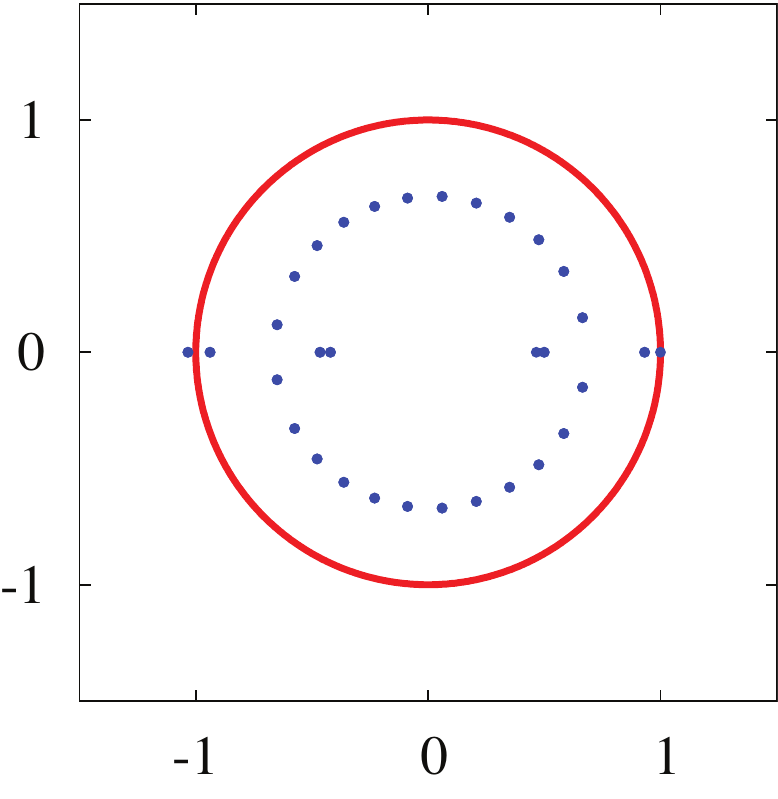}
\caption{Spectral plots in the complex plane for a Wilson-Cowan ring network, with spatial scales $\sigma_{\alpha \beta}=\sigma$ for all $\alpha,\beta$, and $N=31$.  Other parameters as in Fig.~\ref{Fig:Orbit}. 
Left: $\sigma=0.15$, and the synchronous solution is predicted to be linearly stable.
Right: $\sigma=0.191$, and the synchronous solution is predicted to be linearly unstable.
\label{Fig:PWLSpectrumN31}
}
\end{center}
\end{figure}

\begin{figure}[htbp]
\begin{center}
\includegraphics[width=10cm]{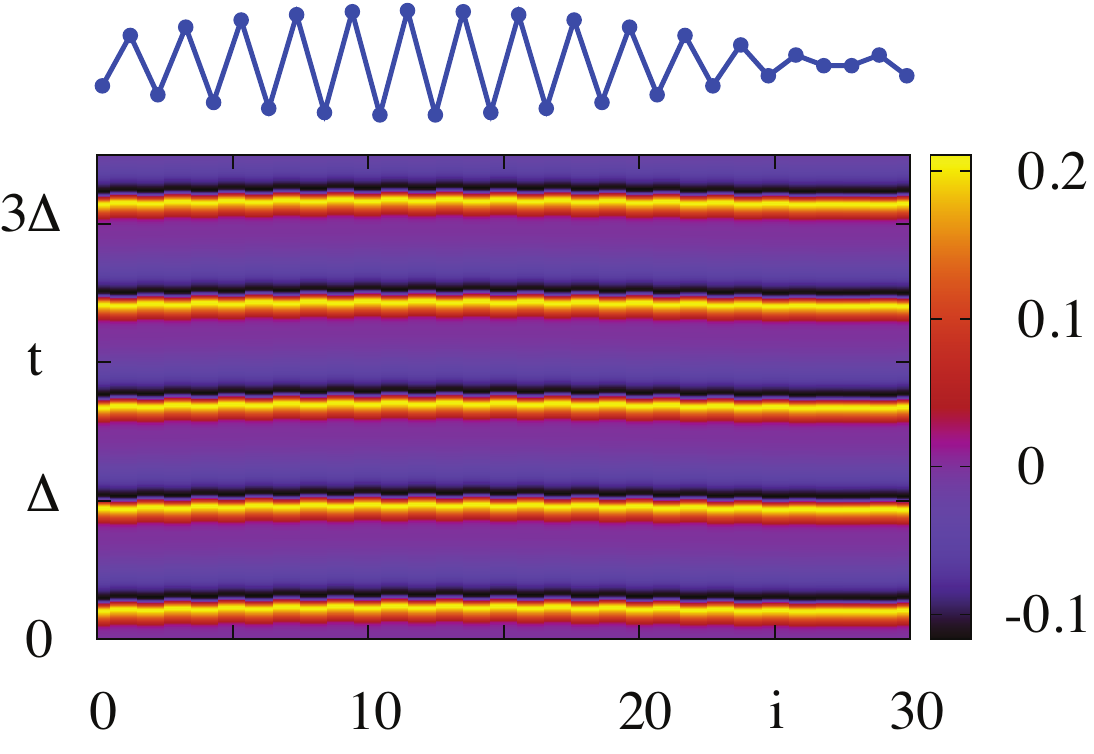}
\caption{Direct numerical simulation of a Wilson-Cowan ring network, with $N=31$, just beyond the point of synchronous instability where $\sigma=0.191$.  Other parameters as in Fig.~\ref{Fig:Orbit}.  Here we plot the components $u_i(t)$ in a space-time plot.  The shape of the unstable mode $e_{p_c}$, with $p_c=16$ (and also $p_c=17$ because of a degeneracy) is depicted in blue at the top of the figure.  The bifurcation point of the linear instability is found to be in excellent agreement with simulations, with the spatial pattern of the emergent network state predicted by $e_{p_c}$.
\label{Fig:PWLWaterfallN31}
}
\end{center}
\end{figure}

By way of illustration of the above theory let us consider a network of Wilson-Cowan nodes arranged on a ring with an odd number of nodes.  Introducing a distance between nodes indexed by $i$ and $j$ as $\operatorname{dist}(i,j)=\min(|i-j|, N-|i-j|)$, we can define a set of exponentially decaying connectivity matrices according to
\begin{equation}
\mathcal{W}^{\alpha \beta}_{ij} = w^{\alpha \beta} \frac{ \e^{-\operatorname{dist}(i,j)/ \sigma_{\alpha \beta}}}{\sum_{j=0}^{N-1} \e^{-\operatorname{dist}(0,j)/\sigma_{\alpha \beta}}} .
\end{equation}
Thus we have a set of four circulant matrices parametrised by the four spatial scales $\sigma_{\alpha \beta}$ that respect the row-sum constraints $\sum_{j=1}^N \mathcal{W}^{\alpha \beta}_{ij} = w^{\alpha \beta}$.
In Fig.~\ref{Fig:PWLSpectrumN31} we show a plot of the eigenvalues of $\Gamma(p)$ for $p=0,\ldots, N-1$ for two different parameter choices.  In one case all of the eigenvalues (excluding the one arising from time-translation invariance) lie within the unit disc, whilst in the other one leaves the unit disc along the negative real axis.  
This latter scenario predicts an instability of the synchronous state, and is consistent with direct numerical simulations.  Moreover, by studying the spectrum under parameter variation we can find the value of $p=p_c$ which goes unstable first.  In Fig.~\ref{Fig:PWLWaterfallN31} we show time courses (obtained by direct numerical simulation) for the components $u_i(t)$ of the emergent network state just beyond the point of instability, as well as a plot of the real part of the spatial eigenvector $e_{p_c}$. We see that the spatial pattern of the network state is well predicted by $e_{p_c}$, suggesting that the bifurcation is supercritical.

\section{\label{sec:Heaviside}The Heaviside world}

In a recent paper Harris and Ermentrout \cite{Harris2015} considered a single Wilson-Cowan population with a Heaviside nonlinearity, where the firing rate in (\ref{WC}) takes the form $F(x)=H(x)$, where $H(x)=0$ for $x<0$ and $H(x)=1$ for $x>0$. 
The choice of a Heaviside firing rate has been very popular in mathematical neuroscience ever since the seminal work of Amari (for neural field models), as nicely exemplified by his recent article on the ``Heaviside World" \cite{Amari13}.  A case in point is the work of Laing and Chow \cite{Laing2002} for understanding binocular rivalry.  They considered a neural mass network model with recurrent excitation, cross-inhibition, adaptation, and synaptic depression and showed that the use of a Heaviside nonlinearity allowed the explicit calculation of the dominance durations of perceptions.  A more recent use of the Heaviside firing rate has been by McCleney and Kilpatrick \cite{McCleney2016} for neural activity models with spike rate adaptation to understand the dynamics of up-down states.  Using techniques from Filippov systems and differential inclusions Harris and Ermentrout made a study of periodic orbits for a Heaviside firing rate using a boundary value problem approach.  Here we show that we can recover their results using the matrix exponential approach of \S \ref{sec:WilsonCowan}.  Moreover, we also extend their work on a single node by showing how to determine the stability of periodic orbits using a nonsmooth version of Floquet theory.  

\begin{figure}[htbp]
\begin{center}
\includegraphics[width=8cm]{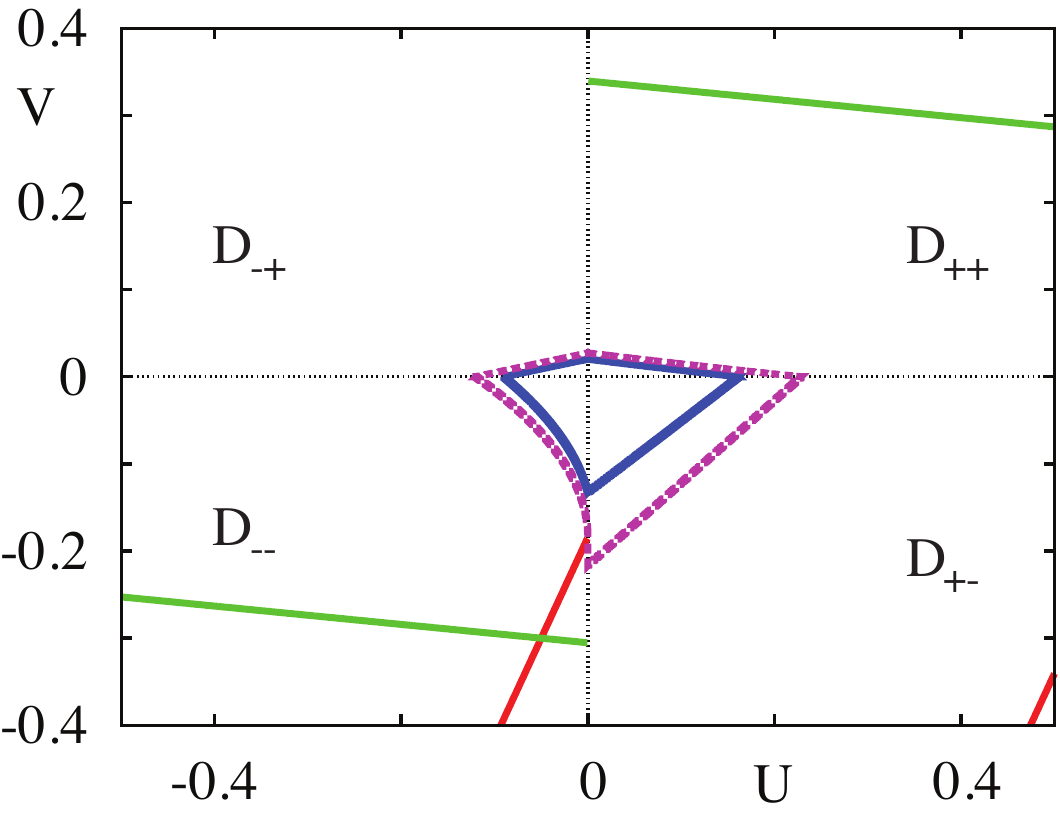}
\caption{Phase plane for a Wilson-Cowan node with a Heaviside firing rate (transformed coordinates), showing the $U$-nullclines (red) and $V$-nullclines (green), as well as a stable periodic orbit (blue), and an unstable periodic sliding orbit (dashed magenta).
Parameters (excluding $\epsilon$) as in Fig.~\ref{Fig:Orbit}.
\label{Fig:WCUV}
}
\end{center}
\end{figure}

In the representation (\ref{WC1}), with $F=H$, we see that the there are two \textit{switching manifolds} defined by $U=0$ and $V=0$.  
If we introduce the indicator functions $h_1(U,V)=U$ and $h_2(U,V)=V$ then we can define these manifolds (lines in this case) as
\begin{equation}
\label{eq:Sigma}
\Sigma_i = \left  \{ (U,V)\in \RSet^2~|~ h_i(U,V)= 0 \right \} .
\end{equation}
These switching manifolds naturally divide the plane into \textit{four} sets.  We denote these by
$D_{++} = \{(U,V)~| U\geq 0, V\geq 0 \}$,
$D_{+-} = \{(U,V)~| U \geq 0, V \leq 0 \}$, $D_{--} = \{(U,V)~| U \leq 0, V \leq 0 \}$, and $D_{-+} = \{(U,V)~| U\leq 0, V \geq 0 \}$. 
If we denote the elements of $A$ by $A_{ij}$, $i=1,2$ and $j=1,2$, 
where
\begin{equation}
A = -\frac{1}{|W|}
\begin{bmatrix}
w^{vu} w^{uv}/\tau - w^{uu} w^{vv} & w^{uu} w^{vu}(1-1/\tau) \\
w^{vv} w^{uv}(1/\tau-1) & w^{uv} w^{vu} - w^{uu} w^{vv}/\tau
\end{bmatrix}
,\qquad |W| = w^{vu} w^{uv} - w^{uu}w^{vv} ,
\end{equation}
then the $U$-nullclines are given by 
\begin{equation}
V=I_v -\frac{A_{11}(U-I_u)}{A_{12}} + \frac{1}{A_{12}}\begin{cases}
-w^{uu} +w^{vu}/\tau & (U,V) \in D_{++} \\
-w^{uu} & (U,V) \in D_{+-}  \\ 
0 & (U,V) \in D_{--} \\ 
w^{vu}/\tau & (U,V) \in D_{-+}
\end{cases}
,
\end{equation}
and the $V$-nullclines are given by
\begin{equation}
V=I_v -\frac{A_{21}(U-I_u)}{A_{22}} + \frac{1}{A_{22}}
\begin{cases}
-w^{uv} +w^{vv}/\tau & (U,V) \in D_{++} \\
-w^{uv} & (U,V) \in D_{+-}  \\ 
0 & (U,V) \in D_{--}  \\ 
w^{vv}/\tau & (U,V) \in D_{-+}
\end{cases}.
\end{equation}
An example set of nullclines is shown in Fig.~\ref{Fig:WCUV}.

To discuss fixed points and their stability it is first necessary to complete the description of the dynamics on the switching manifolds.  
We do this using the \textit{Filippov convex method} \cite{Filippov1998} and extend our 
discontinuous system into a convex differential inclusion.  The Filippov extension of (\ref{WC1}) is then
\begin{equation}
\FD{}{t} \begin{bmatrix} U \\ V \end{bmatrix} \in F(U,V) = 
\begin{cases}
F_{++}(U,V) & (U,V) \in D_{++} \\
\overline{\text{co}} \left ( \{ F_{++}, F_{+-} \} , \kappa_{1}\right ) & (U,V) \in D_{++} \cap D_{+-} \\
F_{+-}(U,V) & (U,V) \in D_{+-} \\
\overline{\text{co}} \left ( \{ F_{+-}, F_{--} \}, \kappa_{2} \right ) & (U,V) \in D_{+-} \cap D_{--} \\
F_{--}(U,V) & (U,V) \in D_{--} \\
\overline{\text{co}} \left ( \{ F_{--}, F_{-+} \}, \kappa_{3} \right ) & (U,V) \in D_{--} \cap D_{-+} \\
F_{-+}(U,V) & (U,V) \in D_{-+} \\
\overline{\text{co}} \left ( \{ F_{-+}, F_{++} \} , \kappa_{4}\right ) & (U,V) \in D_{-+} \cap D_{++} 
\end{cases},
\end{equation}
where $F_{\alpha \beta}(U,V) = A[U-I_u,V-I_v]^T + b_{\alpha \beta}$ for $\alpha, \beta \in \{+,-\}$ and
\begin{equation}
b_{++}=\begin{bmatrix}
w^{uu}-w^{vu}/\tau \\
w^{uv} -w^{vv}/\tau
\end{bmatrix}, \quad
b_{+-}=\begin{bmatrix}
w^{uu}\\
w^{uv}
\end{bmatrix}, \quad
b_{--}=\begin{bmatrix}
0\\
0
\end{bmatrix}, \quad
b_{-+}=\begin{bmatrix}
-w^{vu}/\tau\\
-w^{vv}/\tau
\end{bmatrix}.
\label{XdotHeav}
\end{equation}
Here $\overline{\text{co}} ( \{f,g \}, \kappa) = \kappa f +(1-\kappa)g$ with $\kappa \in [0,1]$ is the closed convex hull of all values between $f$ and $g$.  A \textit{sliding} solution may exist along a switching manifold such that $\dot{h}_i = \nabla h_i \cdot F = 0$.  The functions $\kappa_j$, $j=1,\ldots, 4$, are chosen to ensure that $\dot{h}_i=0$ along any switching manifold.
For example if a sliding solution exists along the line $U=0$ for $V<0$ then we would construct $\kappa_2$ using $\nabla h_1  = (1,0)$ and $F(0,V)=\kappa_2 F_{+-}(0,V)+(1-\kappa_2)F_{--}(0,V)$ yielding
\begin{equation}
\kappa_2 = \frac{(1,0) \cdot  F_{--}(0,V) }{ (1,0) \cdot (F_{--}(0,V)-F_{+-}(0,V)) } .
\label{Filippov}
\end{equation}

As illustrated in Fig.~\ref{Fig:WCUV} it is possible for two nullclines to intersect and create a fixed point $(U_{ss},V_{ss})$.  In the example shown this occurs for $U<0$ and $V<0$, so that $(U_{ss},V_{ss})= (I_u,I_v)$.  Linear stability analysis shows that this is a stable node (with eigenvalues of $A$, namely $-1$ and $-1/\tau$).
Moreover, this system also supports \textit{pseudo} equilibria where either a nullcline touches a switching manifold, or two switching manifolds intersect.  A thorough exploration of the pseudo equilibria of (\ref{WC}) can be found in \cite{Harris2015}.  Here we shall simply focus on the pseudo equilibrium at $(U_{ss},V_{ss})=(0,0)$, and characterise its stability by considering trajectories around this point.  In fact given the PWL nature of the dynamics it is sensible to consider the construction of periodic orbits, and determine the stability of the pseudo equilibrium in terms of the stability of encircling small amplitude orbits.

\subsection{Periodic orbits and their stability}
\label{Sec:stability}

A non-sliding periodic orbit around $(0,0)$ can be constructed in terms of the \textit{times-of-flight} in each region $D_{\alpha \beta}$.  If we denote these four times by the symbols $\Delta_{\alpha \beta}$ then the period of the orbit is given by $\Delta = \Delta_{++}+\Delta_{-+}+\Delta_{--}+\Delta_{+-}$. We may then use a matrix exponential solution:
\begin{equation}
\begin{bmatrix}
U(t) \\ V(t)
\end{bmatrix}
=
\e^{At} 
\begin{bmatrix}
U(0) \\ V(0)
\end{bmatrix}
+(I_2 - \e^{A t}) \left [
\begin{bmatrix}
I_u \\ I_v
\end{bmatrix}
- A^{-1} W J \begin{bmatrix}
H(U) \\ H(V)
\end{bmatrix}
\right ] , \qquad t \geq 0 .
\label{solu}
\end{equation}
to \textit{patch} together solutions, setting the origin of time in each region such that \textit{initial} data in one region comes from \textit{final} data from a trajectory in a neighbouring region.  
We shall denote the periodic orbit by $(\overline{U},\overline{V})$ such that $(\overline{U}(t),\overline{V}(t))=(\overline{U}(t+\Delta),\overline{V}(t+\Delta))$.To indicate which region we are considering we shall simply add $\alpha \beta$ subscripts to the formula in (\ref{solu}).
In this way a periodic orbit that visits all four regions in turn can be parameterised by the five unknowns $\overline{U}_{++}(0)$, $\overline{V}_{++}(\Delta_{++})$, $\overline{U}_{-+}(\Delta_{-+})$, $\overline{V}_{--}(\Delta_{--})$, $\overline{U}_{+-}(\Delta_{+-})$, and $\Delta_{\alpha \beta}$.  These are determined self-consistently by the five equations $\overline{U}_{++}(\Delta_{++})=0$, $\overline{V}_{-+}(\Delta_{-+})=0$, $\overline{U}_{--}(\Delta_{--})=0$, $\overline{V}_{+-}(\Delta_{+-})=0$, and $\overline{U}_{+-}(\Delta_{+-})=\overline{U}_{++}(0)$.  To determine the stability of such an orbit we may use the nonsmooth Floquet theory described in \cite{Coombes2016}.  In essence this treats the propagation of perturbations through a switching manifold using a \textit{saltation} matrix, such that $Y(T^+)= \lim_{\epsilon \searrow 0} Y(T+\epsilon)=
K Y(T)$, where $Y=(U,V)$ denotes the vector state of the system and $K \in \RSet^{2 \times 2}$ is the saltation matrix that acts at time $T$.  Saltation matrices can be derived in a number of ways, with a general prescription in terms of an indicator function $h$ as \cite{Leine2004}
\begin{equation}
K = I_2 +   \frac{\left [ \dot{Y}(T^+) - \dot{Y}(T) \right ] \left [ \vphantom{\dot{Y}(T)} \nabla_Y h(Y(T)) \right ]^\mathsf{T}}{\nabla_Y h(Y(T)) \cdot \dot{Y} (T)} .
\label{K}
\end{equation}
Alternatively, in the context of the PWL model discussed in \S \ref{sec:WilsonCowan}, we can obtain the relevant saltation matrices by considering the approximation $H(x)= \lim_{\epsilon \rightarrow 0} F(x)$.  To see this 
we introduce the vector $\overline{Y}(t) = (\overline{U}(t),\overline{V}(t))$ and linearise the equations of motion (\ref{WC1}) by considering $Y(t) = \overline{Y}(t) + \delta Y(t)$, for small perturbations $\delta Y (t) = (\delta U, \delta V)$.  The linearised equations of motion are given by
\begin{equation}
\FD{}{t} \delta Y = 
\left [A + W J DF(\overline{Y}(t)) \right ] \delta Y .
\label{variation}
\end{equation}
Here $DF(\overline{Y}(t))$ is the piecewise constant matrix described after (\ref{deltaZ}).
Consider for example the time of flight, $t_1(\epsilon)$, between $U=\epsilon$ and $U=0$.  For small $\epsilon$ we may estimate $t_1(\epsilon)$ using the result that $U(t) \simeq U(t_0) + \left. \dot{U} \right |_{t=t_0} (t-t_0)$, giving $t_1(\epsilon)= -\epsilon/ \left. \dot{U} \right |_{t=\Delta_{++}}$.  The corresponding change in state across this small time interval can be obtained by integrating (\ref{variation}) to give 
\begin{equation}
\delta Y(T^+)-\delta Y(T) = \lim_{\epsilon \rightarrow 0}  \int_{T}^{T + t_1(\epsilon)}  W J \begin{bmatrix}
\epsilon^{-1} & 0 \\
0 & 0
\end{bmatrix} \delta Y(t) \d t .
\end{equation}
Thus we obtain 
$\delta Y(T^+) = K_1 \delta Y^-$, with the saltation matrix $K_1$ given by
\begin{equation}
K_1 = I_2 - \frac{1}{\dot{U} (\Delta_{++})} 
WJ
\begin{bmatrix}
1 & 0 \\
0 & 0
\end{bmatrix} .
\label{K1}
\end{equation}
The other saltation matrices (describing the passage through $\epsilon$-neighbourhoods of $U=0$ and $V=0$) are constructed in a similar fashion, and found to be
\begin{align}
K_2&=I_2 - \frac{1}{\dot{V} (\Delta_{-+})} 
WJ
\begin{bmatrix}
0 & 0 \\
0 & 1
\end{bmatrix} ,\nonumber\\
K_3 &= I_2 + \frac{1}{\dot{U} (\Delta_{--})} 
WJ
\begin{bmatrix}
1 & 0 \\
0 & 0
\end{bmatrix} ,\nonumber \\
K_4 &= I_2 + \frac{1}{\dot{V} (\Delta_{+-})} 
WJ
\begin{bmatrix}
0 & 0 \\
0 & 1
\end{bmatrix} .
\label{K234}
\end{align}
It is straightforward to check that the saltation matrices (\ref{K1})-(\ref{K234}) are equivalent to those defined by (\ref{K}).  We now pass to the limit $\epsilon=0$, to treat the Heaviside firing rate.
Between switching events the perturbations evolve according to $\exp(A(t-T)) \delta Y(T^+)$, for $t > T$, where $\delta Y(T^+)$ is the perturbation at the switching time.  Thus after one period of oscillation we may put this all together to obtain
\begin{equation}
\delta Y(\Delta) = \Gamma \delta Y(0), \qquad \Gamma = K_4\e^{A \Delta_{+-}}K_3\e^{A \Delta_{--}}K_2\e^{A \Delta_{-+}}K_1\e^{A \Delta_{++}}.
\end{equation}
The periodic orbit will be stable if the eigenvalues of $\Gamma$ lie within the unit disc.
Note that one of the Floquet multipliers is equal to one, corresponding to perturbations along the periodic orbit.
Let us denote the other eigenvalue by $\e^{\sigma \Delta}$ and use the result that $\det \Gamma = \e^{\sigma \Delta} \times 1$.
Hence,
\begin{align}
\e^{\sigma \Delta} &= \left (\prod_{i=1}^4 \det K_i \right ) \det \e^{A \Delta_{+-}} \det \e^{A \Delta_{--}} \det \e^{A \Delta_{-+}} \det \e^{A \Delta_{++}} \nonumber \\
&= \frac{\dot{\overline{V}}(\Delta_{+-}^+)}{\dot{\overline{V}}(\Delta_{+-})}
\frac{\dot{\overline{U}}(\Delta_{--}^+)}{\dot{\overline{U}}(\Delta_{--})}
\frac{\dot{\overline{V}}(\Delta_{-+}^+)}{\dot{\overline{V}}(\Delta_{-+})}
\frac{\dot{\overline{U}}(\Delta_{++}^+)}{\dot{\overline{U}}(\Delta_{++})} 
\det \e^{A \Delta_{+-}} \det \e^{A \Delta_{--}} \det \e^{A \Delta_{-+}} \det \e^{A \Delta_{++}}. 
\end{align}
Using the fact that $\det \e^{At} = \e^{\operatorname{Tr} A \,t}$ we find
\begin{equation}
\sigma = - \left ( 1+ \frac{1}{\tau} \right ) +\frac{1}{\Delta} 
\log 
\frac{\dot{\overline{V}}(\Delta_{+-}^+)}{\dot{\overline{V}}(\Delta_{+-})}
\frac{\dot{\overline{U}}(\Delta_{--}^+)}{\dot{\overline{U}}(\Delta_{--})}
\frac{\dot{\overline{V}}(\Delta_{-+}^+)}{\dot{\overline{V}}(\Delta_{-+})}
\frac{\dot{\overline{U}}(\Delta_{++}^+)}{\dot{\overline{U}}(\Delta_{++})} .
\label{sigmaHeav}
\end{equation}
A periodic orbit will be stable provided $\sigma <0$.  We shall say that the pseudo-equilibria at $(0,0)$ is unstable (stable) if it is enclosed by a stable (unstable) periodic orbit of arbitrarily small amplitude.  We shall say that there is a \textit{pseudo-Hopf} bifurcation at $(0,0)$ when the pseudo-equilibrium changes stability, namely when $\sigma=0$.  A plot of $\sigma=\sigma(\tau)$ (not shown) for the parameters of Fig.~\ref{Fig:sigma}, shows very similar behaviour as for the steep PWL firing rate function.  In essence we may regard the second term on the right hand side of (\ref{sigmaHeav}) as a correction term to standard Floquet theory to cope with the nonsmooth nature of the Heaviside firing rate.

\subsection{An unstable periodic sliding orbit}

The Wilson-Cowan node can also support an unstable periodic orbit that has a component which slides along the switching manifold $U=0$ for $V \in [V_1,V_2]$, as depicted in Fig.~\ref{Fig:WCUV}.  The points $V_{1,2}$ are easily calculated by determining the points at which the $U$-nullclines touch the switching manifold where $U=0$, and are found to be $V_1=(A_{11}I_u + A_{12}I_V - w^{uu})/A_{12}$ and $V_2= V_1 +w^{uu}/A_{12}$.  In reverse time initial data close to a sliding trajectory would be attracted to it.  Thus we can think of constructing an unstable periodoc sliding orbit, of the type shown in Fig.~\ref{Fig:WCUV}, by breaking it into five pieces.  All pieces of this orbit are constructed similarly to before (see above), except the component that slides.  Using the Filippov method and equation (\ref{Filippov}) we find $\kappa_2 = (A_{11}I_u -A_{12}V +A_{12} I_v)/w^{uu}$, with the sliding dynamics prescribed by 
\begin{equation}
\FD{}{t} \begin{bmatrix} U \\ V \end{bmatrix} = \begin{bmatrix}
0 & 0 \\ 0 & A_{22}-A_{11}w^{uv}/w^{uu}
\end{bmatrix}
\begin{bmatrix} U \\ V  \end{bmatrix} + \begin{bmatrix} 0 \\ b_s \end{bmatrix},
\end{equation}
where $b_s=-A_{12}I_u-A_{22}I_v +(A_{11}I_u+A_{12}I_v)w^{uv}/w^{uu}$.  In backward time the periodic sliding orbit shown in Fig.~\ref{Fig:WCUV} would slide up along $U=0$ until the point $V=V_2$, where it would leave the switching manifold.

We now turn our attention to networks built from Wilson-Cowan nodes with a Heaviside firing rate.

\section{A network of Heaviside Wilson-Cowan nodes}
\label{sec:network_Heaviside}

As we have shown in \S \ref{sec:Heaviside} the replacement of a sigmoidal firing rate by a Heaviside function can lead to highly tractable models for which substantial analytical results can be obtained (with the use of matrix exponentials and saltation matrices).  However, at the network level the mathematical differences between the treatment of smooth and nonsmooth firing rates are considerably amplified relative to those at the single node level.  At the node level it is well known that regarding the Heaviside function as the steep limit of a sigmoidal function can lead to arbitrarily many different non-equivalent dynamical systems. This is simply due to the non-uniqueness of the singular limits by which smooth functions may tend towards discontinuities.  For a recent perspective on this issue see the work of Jeffrey \cite{Jeffrey2015}.
Thus there is no reason to assume that taking the limit $\epsilon \rightarrow 0$ for the PWL network considered in \S \ref{sec:WilsonCowanNetwork} will be relevant to a Wilson-Cowan network with a Heaviside nonlinearity.
Namely the approximation of a Heaviside function by a continuous function such that $H(x) = \lim_{\epsilon \rightarrow 0} F(x)$, where  $F(x)$ is given by (\ref{pwlrate}), may have little utility given that pointwise convergence need not imply distributional convergence.

We now return to the network introduced in \S \ref{sec:WilsonCowanNetwork}, but replace the dynamics of each node with the Heaviside limit studied in the previous section. For the following analysis, it is convenient to rewrite \eqref{eq:WCN_dY} as
\begin{equation}
\label{eq:WCN_Heavi_dY}
\FD{}{t} Y = \mathcal{A} (Y- \mathcal{F}(Y))\,,\qquad \mathcal{F}(Y)= C-\mathcal{A}^{-1}\mathcal{W} \mathcal{J} H(Y).
\end{equation}
The network model (\ref{eq:WCN_dY}), with a Heaviside nonlinearity, is reminiscent of a so-called \textit{Glass} network originally introduced for the study of biochemical networks that are dominated by switch-like behaviour \cite{Glass1973,Glass1975}, though here the model has two-time scales.  For a nice survey of periodic and aperiodic behaviour in Glass networks we recommend the article by Edwards \cite{Edwards2000}, and for the application to gene networks see Edwards and Glass \cite{Edwards2000a}.

The synchronous network state is given by (\ref{solu})  (remembering the row-sum constraint on the network connections).  To study its linear stability we consider values of the perturbed network state $Y$ that are close to the synchronous network state at the unperturbed crossing times. Let $\overline T_i$ denote the time that the synchronous state moves between one of the four quadrants (as illustated in Fig.~\ref{Fig:WCUV}). We then make the ansatz that the perturbed network state $Y$ can be expressed with respect to the synchronous orbit at one of the switching times $\overline T_i$ and write $Y(t)=\overline Y(\overline T_i)+\delta Y(t)$ with $t$ in the neighbourhood of $\overline T_i$. 

We first construct the saltation matrix through a switch, indexed by $i=1,\ldots,4$.  Suppose that the $k$th crossing occurs at a perturbed crossing time $T_{i,k}$. The network states at two consecutive crossings are related via
\begin{equation}
\label{eq:Y_kplus1_minus}
Y(T_{i,k+1})=\e^{ \mathcal{A} (T_{i,k+1} - T_{i,k})} Y(T_{i,k})+\left(I_{2N}- \e^{ \mathcal{A} (T_{i,k+1} - T_{i,k}) }\right) \mathcal{F}(Y(T_{i,k}^+))\, .
\end{equation}
This equation is obtained by integrating (\ref{eq:WCN_Heavi_dY}) using the observation that $\mathcal{F}$ is constant between crossings.
By linearising (\ref{eq:Y_kplus1_minus}) we can relate the perturbations between crossing events as
\begin{equation}
\label{eq:deltaY_ikplus1}
\delta Y(T_{i,k+1})=\delta Y(T_{i,k})+Y^{i,k} \delta T_{i,k}\,,
\end{equation}
where $Y^{i,k}=\mathcal A(\overline Y(\overline T_i)- \mathcal{F}(Y(T_{i,k}^+)))$ and $\delta T_{i,k}=T_{i,k+1}-T_{i,k}$. For the node that crosses at $T_{i,k+1}$, the corresponding component of $\delta Y(T_{i,k+1})$, say at position $m$, vanishes, since $Y_m(T_{i,k+1})=\overline Y_m(\overline T_i)$ (namely the $m$th component of the perturbed trajectory equals the $m$th component of the synchronous orbit). Here, $m \in \{1,3,\ldots, 2N-1\}$ or $m\in\{2,4,\ldots 2N\}$, depending on whether the crossing occurs along the $V$ or $U$ axis. We then see from \eqref{eq:deltaY_ikplus1} that
\begin{equation}
\label{eq:deltaT_ik}
\delta T_{i,k}=-\frac{\delta Y_m(T_{i,k})}{Y^{i,k}_m} \,.
\end{equation}
At this point, $m$ is still unknown. However, since $m$ corresponds to the node that crosses before any of the other remaining nodes do so, we find it by minimising \eqref{eq:deltaT_ik} over the possible values of $m$, and we denote it by $m_k$. When we combine \eqref{eq:deltaY_ikplus1} and \eqref{eq:deltaT_ik}, we find that $\delta Y(T_{i,k+1})=\Gamma_{i,k} \delta Y(T_{i,k})$ with
\begin{equation}
\label{eq:Gamma_ik}
\Gamma_{i,k}=I_{2N}-\frac{Y^{i,k} e_{m_k}^\mathsf{T}}{Y^{i,k}_{m_k}}\\,
\end{equation}
where $e_m$ is the $m$th canonical basis vector in $\mathbb R^{2N}$. The saltation matrix for each of the four switches is then given by
\begin{equation}
L_i=\Gamma_{i,N-1} \Gamma_{i,N-2} \cdots \Gamma_{i,1}\,, \qquad i=1,\ldots, 4.
\label{Li}
\end{equation}
The ordering of matrix multiplications in (\ref{Li}) is determined by the iterative minimisation of the perturbations given by (\ref{eq:deltaT_ik}).

In the next step, we analyse how a perturbed network state is propagated between saltation events. Let $T_i^+$ denote the time when the last node crosses between quadrants. Here, the superscript makes explicit that all nodes have crossed into the next quadrant. The next network event occurs when one of the nodes crosses into the subsequent quadrant.  This happens at a time $T_{i+1}^-$, where the superscript indicates that only one node has crossed. We will make the ansatz that $T_i^+=\overline T_i +\delta T_i^+$ and $T_{i+1}^-=\overline T_{i+1} +\delta T_{i+1}^-$. We see from \eqref{eq:WCN_Heavi_dY} that
\begin{equation}
\label{eq:Y_iplus1_minus}
Y(T_{i+1}^-)=\e^{ \mathcal{A} (T_{i+1}^- - T_i^+)} Y(T_i^+)+\left(I_{2N}- \e^{ \mathcal{A} (T_{i+1}^- - T_i^+) }\right) \mathcal{F}(Y(T_i^+))\,,
\end{equation}
from which we obtain after linearisation
\begin{equation}
\label{eq:deltaY_iplus1_minus}
\delta Y(T_{i+1}^-)=\e^{\mathcal A \Delta_i}\left( \delta Y(T_i^+) - \overline{Y}'(\overline T_i^+) \delta T_i^+\right)+\overline{Y}'(\overline T_{i+1}^-) \delta T_{i+1}^-\,,\\
\end{equation}
where we have used the fact that $\mathcal{F}(Y(T_i^+))=\mathcal{F}(Y(T_{i+1}^-))$, since $\mathcal{F}$ is constant between crossing events. Here $\overline{Y}'(t)$ denotes the differential of $\overline{Y}(t)$ with respect to $t$.
As above, the component of $\delta Y(T_{i+1}^-)$ that corresponds to the node that switches first, say at position $m$, vanishes. Taking the $m$th component of \eqref{eq:deltaY_iplus1_minus} then yields an expression for the perturbation of the crossing time
\begin{equation}
\label{eq:deltaT_iplus1_minus}
\delta T_{i+1}^-=-\frac{f_m^i}{\overline{Y}_m'(\overline T_{i+1}^-)}\,,
\end{equation}
where the vector $f^i \in \RSet^{2N}$ is given by $\e^{\mathcal A \Delta_i}\left( \delta Y(T_i^+) - \overline{Y}'(\overline T_i^+) \delta T_i^+\right)$. We again find the value of $m$ by minimising \eqref{eq:deltaT_iplus1_minus} over all admissible values of $m$ and refer to it as $m_i$. This leads to $\delta Y(T_{i+1}^-)=\Gamma_i \delta Y(T_i^+)$ with
\begin{equation}
\Gamma_i=\left(G_i-\frac{\overline Y'(\overline T_{i+1}^-)}{\overline{Y}_{m_i}'(\overline T_{i+1}^-)} e_{m_i}^T G_i \right)\,,\qquad
\end{equation}
and
\begin{equation}
G_i=\e^{\mathcal A \Delta_i}\left( I_{2N} - \frac{\overline{Y}'(\overline T_i^+) e_1^T\delta T_i^+}{\delta Y_1(T_i^+)}\right) \,.
\end{equation}
Taken together, we obtain after one period 
\begin{equation}
\delta Y(T_4^+)=\Psi \delta Y(0)\,, \qquad \Psi=L_4 \Gamma_4 L_3 \Gamma_3 L_2 \Gamma_2 L_1 \Gamma_1\,.
\end{equation}
The matrices $\Gamma_i$ act to propagate perturbations across a quadrant, and the $L_i$ propagate perturbations through a switch.
At first sight, the definition of $G_i$ suggests that we have introduced a dependence of $\Gamma_i$ on $\delta Y(0)$ through the inclusion of $\delta Y(T_i^+)$. This dependence can be avoided by noting that $\delta T_i^+=\delta T_i^-+\sum_k \delta T_{i,k}$ and the repeated use of \eqref{eq:deltaT_ik}, \eqref{eq:Gamma_ik} and \eqref{eq:deltaT_iplus1_minus}. The drawback of this approach is that the resultant operator does not lend itself to an interpretation of successive propagations and saltations, nor is it numerically advantageous. Moreover, this operator would only remove the \emph{explicit} dependence of $\Psi$ on $\delta Y(0)$. The minimisation steps that are necessary to determine the order in which nodes switch already leads to an \emph{implicit} dependence of $\Psi$ on $\delta Y(0)$. Changing $\delta Y(0)$ can lead to a different order of switching, and since matrix multiplication does not commute, $\Psi$ can be different for different $\delta Y(0)$.
This has profound implications for asserting linear stability. The usual argument that the eigenvalues of $\Psi$ determine linear stability does not hold anymore. To see this, consider the propagation of $\delta Y(0)$ over multiple periods, i.e.
\begin{equation}
\delta Y^{(1)}=\Psi^{(0)} \delta Y(0)\,,\quad \delta Y^{(2)}=\Psi^{(1)} \delta Y^{(1)}\,,\quad \delta Y^{(3)}=\Psi^{(2)} \delta Y^{(2)}\,,\quad \ldots
\end{equation}
so that
\begin{equation}
\label{eq:deltaY_m}
\delta Y^{(m)}=\Psi^{(m-1)} \Psi^{(m-2)}\cdots\Psi^{(0)}\delta Y(0)\,.
\end{equation}
The eigenvalues of $\Psi^{(i)}$ and $\Psi^{(j)}$ can be different for $i \neq j$.  For some value of $i$ $\Psi^{(i)}$ may have all eigenvalues in the unit disc, whilst for another value of $i$ there may be some eigenvalues outside the unit disc. Over one period, perturbations can therefore grow or shrink. This entails that for a product of operators as in \eqref{eq:deltaY_m}, $\delta Y^{(m)}$ may be smaller than $\delta Y^{(0)}$, although some $\Psi^{(i)}$ might have some eigenvalues that lie outside the unit disc.  Instead of looking at the eigenvalues of individual $\Psi^{(i)}$, we could have studied the eigenvalues of the product of operators in \eqref{eq:deltaY_m}. We would have come to the same conclusion since eigenvalues of the product operator move into and out of the unit disc as we increase $m$.

\begin{figure}[htbp]
\begin{center}
\includegraphics[height=4cm]{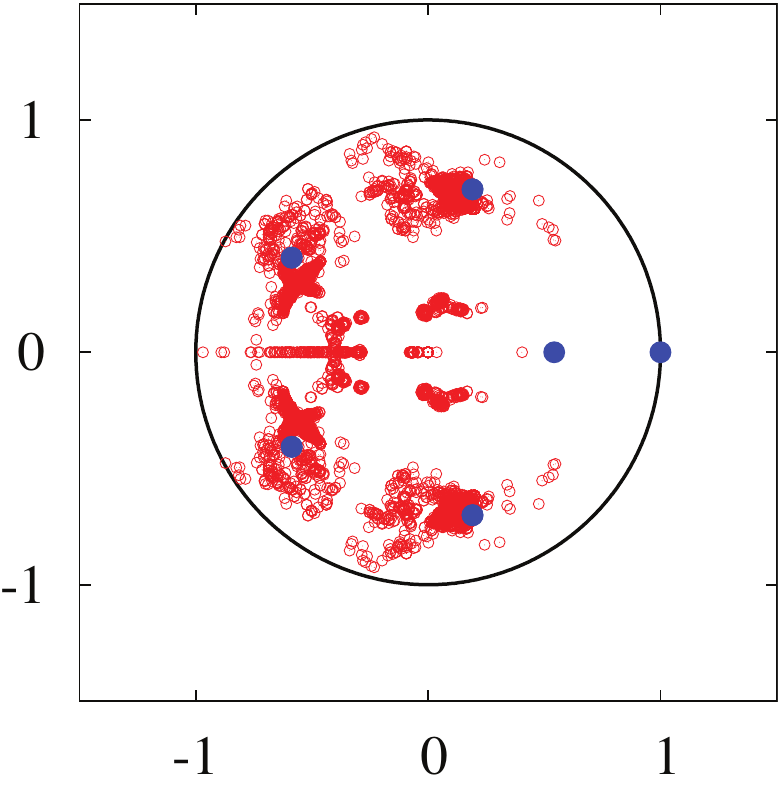} \hfill
\includegraphics[height=4cm]{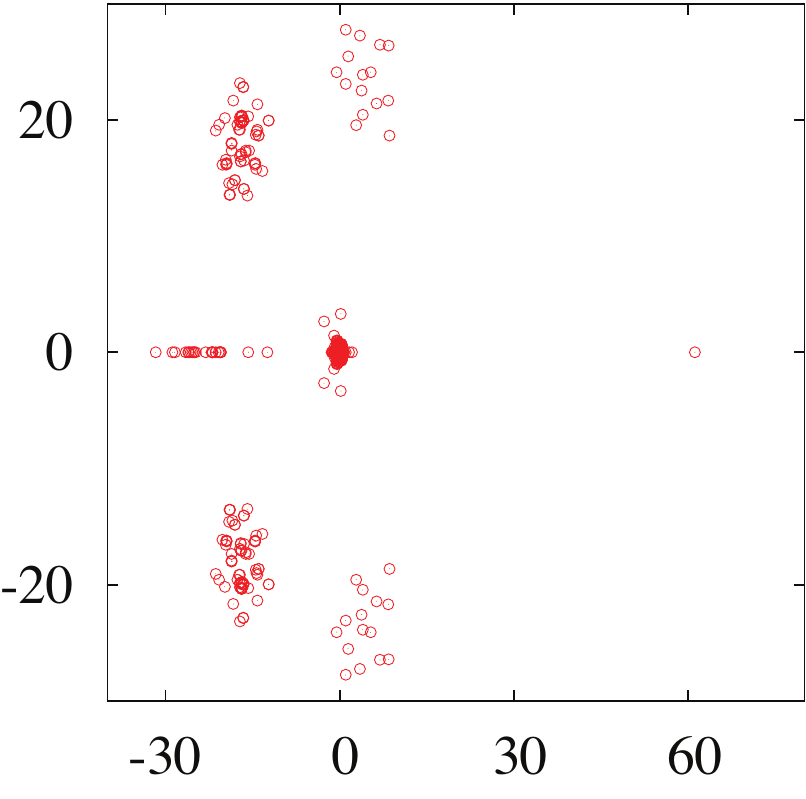} \hfill
\includegraphics[height=4cm]{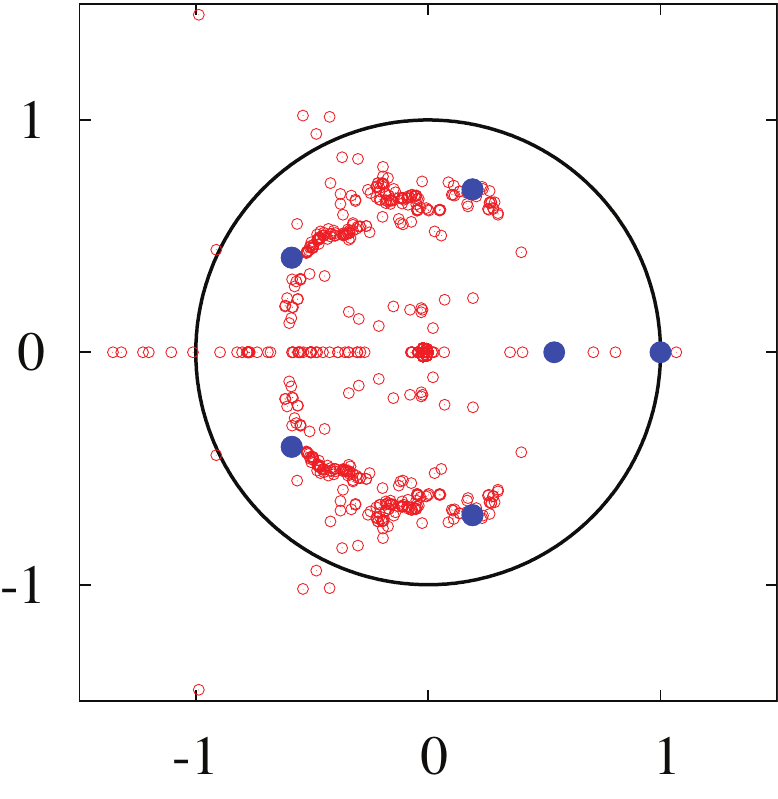}
\caption{Spectral plots for a Heaviside Wilson-Cowan ring network with spatial scales $\sigma_{\alpha \beta}=0.215$ for all $\alpha,\beta$, and $N=5$. We sampled $2000$ random initial conditions, and eigenvalues are shown as open red circles. The filled blue circles are the eigenvalues of the PWL network with the same parameter values and $\epsilon=0.001$. (Left) Spectra for initial conditions that lead to eigenvalues that all fall into the unit disc. (Middle) Spectra for initial conditions that lead to eigenvalues outside the unit disc. (Right) Blow-up of the the middle panel around the unit disc.
Other parameters as in Fig.~\ref{Fig:Orbit}.
\label{Fig:Spectra-N5-sigma_0d215}
}
\end{center}
\end{figure}

\begin{figure}[htbp]
\begin{center}
\includegraphics[height=4cm]{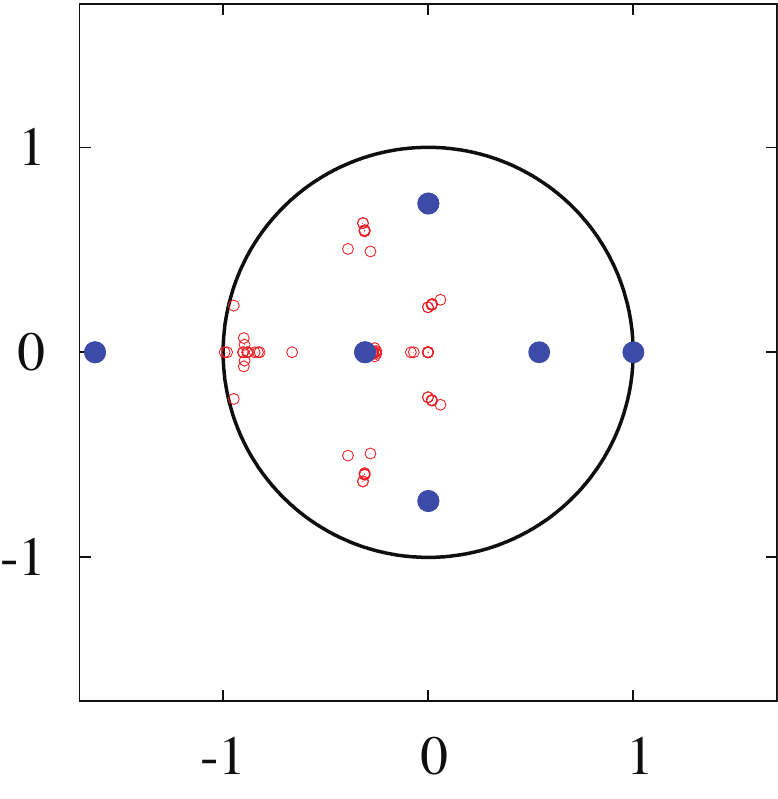} \hfill
\includegraphics[height=4cm]{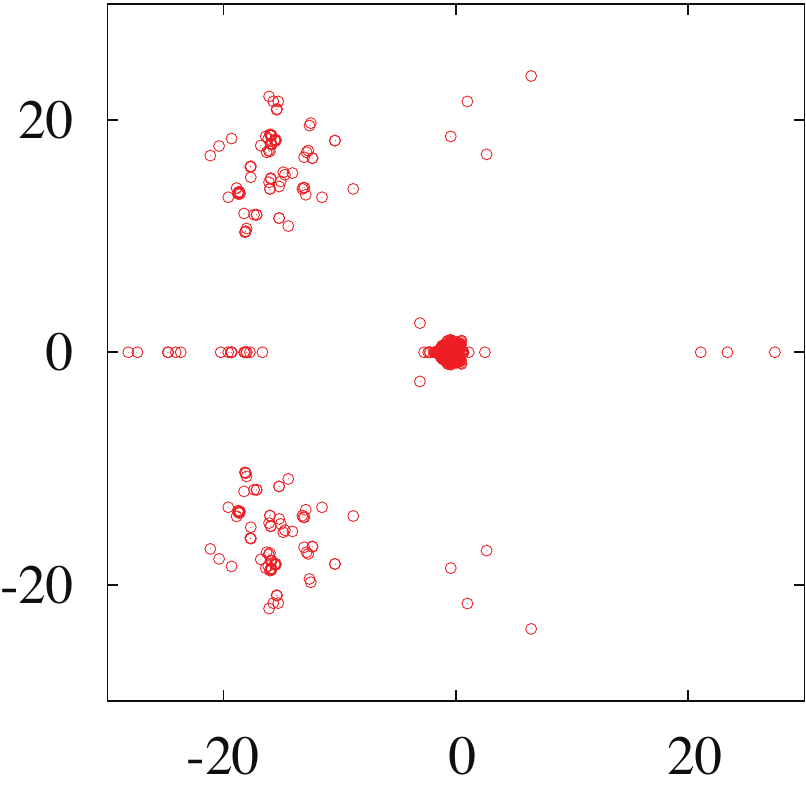} \hfill
\includegraphics[height=4cm]{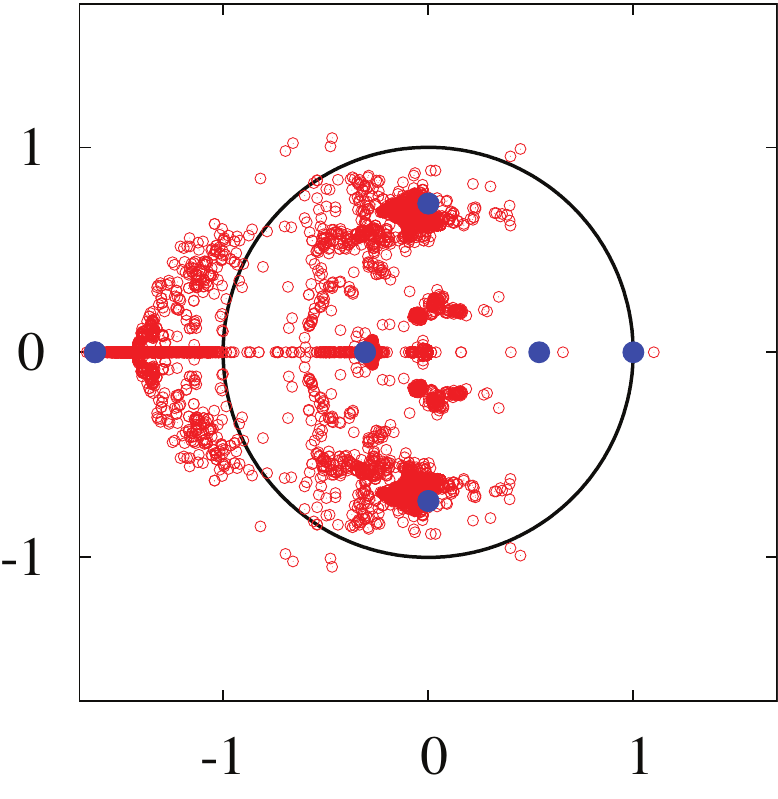}
\caption{Spectral plots for a Heaviside Wilson-Cowan ring network with spatial scales $\sigma_{\alpha \beta}=0.23$ for all $\alpha,\beta$, and $N=5$. We sampled $2000$ random initial conditions, and eigenvalues are shown as open red circles. The filled blue circles are the eigenvalues of the PWL network with the same parameter values with $\epsilon=0.001$. (Left) Spectra for initial conditions that lead to eigenvalues that all fall into the unit disc. (Middle) Spectra for initial conditions that lead to eigenvalues outside the unit disc. (Right) Blow-up of the the middle panel around the unit disc.
Other parameters as in Fig.~\ref{Fig:Orbit}.
\label{Fig:Spectra-N5-sigma_0d23}
}
\end{center}
\end{figure}

Figures \ref{Fig:Spectra-N5-sigma_0d215} and \ref{Fig:Spectra-N5-sigma_0d23} illustrate the dependence of the spectra on random initial conditions $\delta Y(0)$. In both figures, the left panel shows the spectra for initial conditions when all eigenvalues of $\Psi^{(0)}$ lie within the unit disc. The middle panel displays spectra with some  eigenvalues outside the unit disc, and the right panel is a blowup of the middle panel around the unit disc. For Fig.~\ref{Fig:Spectra-N5-sigma_0d215}, we chose a value of $\sigma$ such that the synchronous orbit of the PWL network, with a small values of $\epsilon=0.001$, is linearly stable. We observe that the eigenvalues of the Heaviside network cluster around those of the PWL network. While it appears that the majority of synchronous solutions are stable (for this parameter choice), some initial conditions lead to eigenvalues outside the unit disc. When zooming into the unit disc, we see some degree of clustering, although this is not as pronounced as for the stable solutions. 

For larger values of $\sigma$, the synchronous state of the PWL network becomes unstable (for small $\epsilon)$. The left panel of Fig.~\ref{Fig:Spectra-N5-sigma_0d23} shows that the eigenvalues of the Heaviside network that all fall into the unit disc exhibit only a weak association with the eigenvalues of the PWL network. In addition, it seems that more initial conditions lead to unstable synchronous solutions than stable ones. This mirrors the behaviour in Fig.~\ref{Fig:Spectra-N5-sigma_0d215}, where the majority of initial conditions gives rise to stable solutions. The blow-up in the right panel of Fig.~\ref{Fig:Spectra-N5-sigma_0d23} illustrates that the eigenvalues of the Heaviside network form clusters around those of the PWL network.
While the notion of linear stability in terms of eigenvalues of the propagator is lost for the Heaviside network, it appears that the clustering of these eigenvalues reflects the stability of the PWL system, at least for small values of $\epsilon$ (where the PWL firing rate becomes more switch like).

\section{\label{sec:conclusion}Conclusion}

In this paper we have shown that the combination of two popular approaches in dynamical systems, namely PWL modelling of low dimensional oscillators and the MSF, can be combined to give insight into the behaviour of network states in neural mass network models.  This is natural for this type of system since the sigmoidal nonlinearity, ubiquitous throughout neuroscience modelling of large scale brain dynamics, is well caricatured by a PWL reduction.
We have focused here on the bifurcation of the synchronous network state, and shown how this can be determined in terms of a set of low-dimensional Floquet problems, each of which can be solved using simple linear algebra.  In essence the PWL aspect of the model allows the variational problem for stability to be solved without recourse to the numerical solution of an ordinary differential equation.  Closed form solutions are \textit{patched} together, and although this may appear inelegant at first sight, it does lead to explicit formulas for Floquet exponents at the single node level, and is easily cast into algorithmic form for accurate numerical computations at the network level.  This nicely highlights the benefits of PWL modelling.  Importantly the approach advocated here is not just limited to the construction and stability of the synchronous state.  Pecora \textit{et al}. \cite{Pecora2014}  and Sorrentino \textit{et al}. \cite{Sorrentino2016} have recently extended the MSF approach to treat more exotic states making extensive use of tools from computational group theory.  Thus the work presented here is readily extended to treat non-synchronous states, such as clusters, and for a further discussion see \cite{Nicks2017}.  From a neuroscience perspective it would also be important to treat delays, arising from the finite propagation speed of action potentials relaying signals between distinct brain regions \cite{Deco09}.  In this case we would hope to exploit the growing body of knowledge on PWL dynamics with time delay, as exemplified by \cite{Senthilkumar2005}.  

From a mathematical perspective we have also seen that there is an important difference between the analysis of a high gain continuous PWL sigmoid and that of a discontinuous switch-like Heaviside firing rate.  Although this can be facilitated with the use of saltation matrices (to propagate perturbations through switching manifolds) there is no MSF style approach that reduces the study of synchrony to a set of sub-network Floquet problems.  Moreover, in contrast to the linear stability analysis of continuous systems, there is now a new challenge of addressing the temporal order in which perturbations to network states pass through a switching manifold.  To treat this we have made use of ideas originally developed for Glass networks \cite{Edwards2000}, though note that similar issues of \textit{ordering} also arise in the analysis of pulse-coupled systems \cite{Timme2002,Kielblock2011,Goel2002}. In essence the analysis of a Wilson-Cowan network with a Heaviside firing rate must be performed carefully, and with non-standard tools, as its behaviour can differ from that of a similar network with a high gain PWL sigmoid.

\begin{acknowledgements}
This work was supported by the Engineering and Physical Sciences Research Council [grant number EP/P007031/1].
M. \c{S}ayli was supported by a grant from T\"UBITAK.
\end{acknowledgements}

\bibliographystyle{unsrt}

\label{lastpage}

\end{document}